\documentclass[aps,twocolumn,pra,showpacs,superscriptaddress,footinbib,tightenlines]{revtex4-2}
\usepackage{amsmath}
\usepackage{graphicx}
\usepackage{color}
\usepackage{amsfonts}
\usepackage[colorlinks,citecolor=blue]{hyperref}
\begin{document}

\title{Deterministic generation of nonclassical mechanical states in cavity optomechanics \\ via reinforcement learning}
\author{Yu-Hong Liu}
\affiliation{Key Laboratory of Low-Dimensional Quantum Structures and Quantum Control of
Ministry of Education, Key Laboratory for Matter Microstructure and Function of Hunan Province, Department of Physics and Synergetic Innovation Center for Quantum Effects and Applications, Hunan Normal University, Changsha, Hunan 410081, China}
\affiliation{Hunan Research Center of the Basic Discipline for Quantum Effects and Quantum Technologies, Hunan Normal University, Changsha, Hunan 410081, China}
\author{Qing-Shou Tan}
\affiliation{Key Laboratory of Hunan Province on Information Photonics and Freespace Optical Communication, College of Physics and Electronics, Hunan Institute of Science and Technology, Yueyang, Hunan 414000, China}
\author{Le-Man Kuang}
\affiliation{Key Laboratory of Low-Dimensional Quantum Structures and Quantum Control of
	Ministry of Education, Key Laboratory for Matter Microstructure and Function of Hunan Province, Department of Physics and Synergetic Innovation Center for Quantum Effects and Applications, Hunan Normal University, Changsha, Hunan 410081, China}
\affiliation{Hunan Research Center of the Basic Discipline for Quantum Effects and Quantum Technologies, Hunan Normal University, Changsha, Hunan 410081, China}
\author{Jie-Qiao Liao}
\email{Contact author: jqliao@hunnu.edu.cn}
\affiliation{Key Laboratory of Low-Dimensional Quantum Structures and Quantum Control of
Ministry of Education, Key Laboratory for Matter Microstructure and Function of Hunan Province, Department of Physics and Synergetic Innovation Center for Quantum Effects and Applications, Hunan Normal University, Changsha, Hunan 410081, China}
\affiliation{Hunan Research Center of the Basic Discipline for Quantum Effects and Quantum Technologies, Hunan Normal University, Changsha, Hunan 410081, China}
\affiliation{Institute of Interdisciplinary Studies, Hunan Normal
University, Changsha, Hunan 410081, China}

\begin{abstract}
Nonclassical mechanical states, as vital quantum resources for exploring macroscopic quantum behavior, have wide applications in the study of the fundamental quantum mechanics and modern quantum technology. In this work, we propose a scheme for deterministically generating non-classical mechanical states in cavity optomechanical systems. By working in the eigen-representation of the nonlinear optomechanical systems, we identify the carrier-wave resonance conditions and seek  optimal driving pulses for state preparations. Concretely, we employ the reinforcement learning method to optimize the pulsed driving fields, effectively suppressing the undesired transitions induced by both the pulsed driving fields and dissipations. This approach enables the high-fidelity preparation of phononic Fock states and superposed Fock states in the single-resonator optomechanical systems, as well as two-mode entangled states in the two-resonator optomechanical systems. The statistical properties of the generated states are also examined. Our results create an opportunity for quantum state engineering in quantum optics and quantum information science via reinforcement learning.

\end{abstract}

\date{\today}
\maketitle

\section{INTRODUCTION}
Cavity optomechanics addresses the understanding, manipulation, and application of the radiation-pressure interaction between macroscopic mechanical resonators and electromagnetic fields in a cavity~\cite{SC2008Kippenberg,AspelmeyerPT2012,AspelmeyerRMP2014}. One of the interesting tasks in cavity optomechanics is the generation of nonclassical states in the macroscopic mechanical degrees of freedom, such as the squeezed states~\cite{MariPRL2009,ZollerPRA2009,NunnenkampPRA2010,Liaopra2011,WollmanSc2015,LecocqPRX2015,PirkkalainenPRL2015}, Fock states~\cite{GallandPRL2014,HongSc2017}, entangled mechanical states~\cite{ManciniPRL2002,Riedinger2018,Ockeloen-Korppi2018,ShlomiSC2021,LaureSc2021}, and cat states~\cite{BosePRA1997,MarshallPRL2003,BassiPRL2005,LiaoPRL2016,HoffPRL2016,JeanniPRL2018,FlorianRMP2018,LI202315}. These macroscopic quantum states have significant applications in the study of fundamental quantum mechanics~\cite{AspelmeyerPT2012}, ultrahigh-precision measurements~\cite{MetcalfeAPR2015,CavesRMP1980}, and quantum information processing~\cite{StannigelPRL2010,FiorePRL2011}.

So far, various methods have been explored to generate  nonclassical mechanical states in cavity optomechanics, including conditional dynamics~\cite{MarshallPRL2003,LiaoPRL2016}, nonlinear coupling~\cite{BosePRA1997,MarshallPRL2003,LiaoPRL2016,XuPRA2013,GarzianoPRA2016,Latmiral2018,HauerPRL2023,WisePRA2024}, state transfer~\cite{HoffPRL2016,KiesewetterPRA2018,JiePRA2020}, photon or phonon subtraction~\cite{PaternostroPRL2011,VannerPRX2011,Brawley2016,ShomroniPRA2020,HuPRR2023}, and reservoir engineering~\cite{TianPRL2013,Ying-DanPRL2013,KienzlerSC2015,YouPRA2015}. In cavity optomechanics, the optical drivings provide a powerful tool for controlling the dynamics of the system~\cite{VannerPnas2011}, and hence the optimization of the optical driving offers a promising route for controlling the cavity optomechanical systems. For instance, the monochromatic strong driving of the optomechanical cavity can lead to the linearization of the systems~\cite{AspelmeyerRMP2014,bowen2015quantum} and further realize the optomechanical cooling~\cite{Wilson-RaePRL2007,MarquardtPRL2007,ChanNature2011,TeufelNature2011,LiuPRA2022}, optomechanical entanglement~\cite{VitaliPRL2007,PalomakiSc2013,Yu2020,JiaoPRL2020,LaiPRL2022,JiaoPRAA2022,LiuYHPRA2024,JiaoLPR2024}, and mechanical squeezing~\cite{MariPRL2009,ZollerPRA2009,Liaopra2011,LiuPRA2024}. In contrast, implementing optimal driving in the single-photon strong-coupling regime~\cite{AspelmeyerRMP2014,bowen2015quantum} is a hard task, because the nonlinear couplings and complex energy-level transitions exist in the optomechanical systems working in this regime. Recently, reinforcement learning has emerged as a powerful method to address these challenges by dynamically optimizing the control strategies based on reward and penalty signals~\cite{Sutton2018,YaoPRX2021,TanPRA2021,ZengPRL2023,ZhangCP2023,ZengPRL2025}. In particular, several recent studies have applied reinforcement learning to cavity optomechanical systems, revealing its potential for quantum control tasks such as motional cooling~\cite{SarmaPRR2022} and entanglement engineering~\cite{YeAPLML2025}. Consequently, reinforcement learning is expected to provide new insights for optimizing the control of cavity optomechanical systems in the single-photon strong-coupling regime, enabling more precise and efficient manipulation of their quantum states and dynamics. Note that mechanical Fock states have been prepared in circuit quantum acoustodynamics using quantum optimal control~\cite{rahman2024genuine}.

In this work, we apply the reinforcement learning method to prepare  nonclassical mechanical states in a cavity optomechanical system operating in the single-photon ultrastrong-coupling regime, namely, the single-photon optomechanical coupling is comparable to the mechanical frequency. Concretely, we consider the generation of phononic Fock states and superposed Fock states in the single-resonator optomechanical systems, as well as two-mechanical-mode entangled states in the two-resonator optomechanical systems. By analyzing the eigensystem of the undriven optomechanical system, we determine the carrier-wave resonance conditions, and further apply reinforcement learning to optimize the driving pulses, ensuring the high-fidelity generation of nonclassical mechanical states. This work demonstrates the application of reinforcement learning in optimal control of cavity optomechanical systems.

The rest of this paper is organized as follows. In Sec.~\ref{Sec2}, we show the generation of  phononic Fock states and superposed Fock states in the single-resonator optomechanical system. In Sec.~\ref{Sec3}, we extend the method to the two-resonator optomechanical system and investigate the preparation of entangle states between the two resonators. We discuss the experimental feasibility of our proposal, compare the reinforcement learning scheme and the stimulated Raman adiabatic passage (STIRAP) approach for state generation, investigate the detection of mechanical nonclassical states, and summarize this work in Sec.~\ref{Sec4}.

\begin{figure}[tbp]
\centering \includegraphics[width=0.45\textwidth]{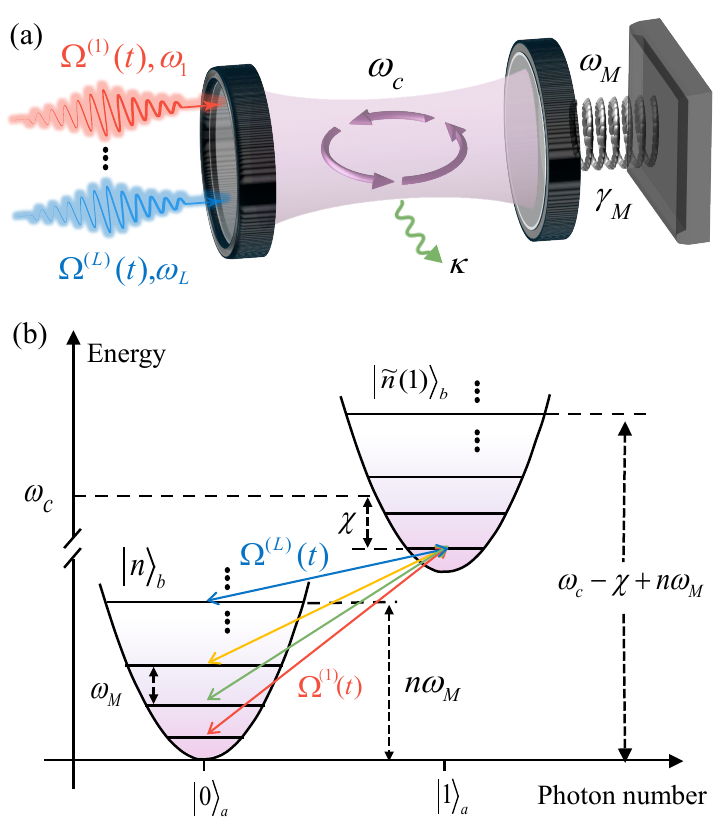}
	\caption{(a) Schematic of a single-resonator optomechanical system, where the optical cavity (with resonance frequency $\omega_{c}$) is driven by $L$ pulsed driving fields $\Omega^{(1)}(t)$, $\Omega^{(2)}(t)$, ..., $\Omega^{(L)}(t)$, with the carrier frequencies $\omega_{1}$, $\omega_{2}$, ..., $\omega_{L}$, respectively. The mechanical mode has a resonance frequency $\omega_{M}$. The decay rates of the cavity field and the mechanical mode are $\kappa$ and $\gamma_{M}$, respectively.  (b) Unscaled energy-level structure of the single-resonator optomechanical system restricted within the zero- and one-photon subspaces. The pulsed driving fields induce transitions between adjacent photon subspaces, enabling controllable quantum state preparation.}
\label{Fig1}
\end{figure}

\section{Generation of phononic Fock states and superposed Fock states in the single-mechanical-resonator optomechanical system\label{Sec2}}
In this section, we explore the generation of phononic Fock states and superposed Fock states in a single-resonator optomechanical system with the application of reinforcement learning. We begin by introducing the physical model and its Hamiltonian, followed by the derivation of the effective Hamiltonian within a restricted Hilbert space. Subsequently, the open-system dynamics of the optomechanical system in the single-photon ultrastrong-coupling regime is analyzed using the dressed master equation. Finally, we employ reinforcement learning, implemented via the deep deterministic policy gradient (DDPG) algorithm~\cite{SumieaHe2024}, to optimize the pulsed driving fields, enabling the high-fidelity preparation of phononic Fock states and superposed Fock states.

\subsection{Physical model and Hamiltonians}
We consider a single-resonator cavity optomechanical system, in which the cavity fields interact with a mechanical resonator via the radiation-pressure coupling. We consider the adiabatic regime of the optomechanical system and then we focus on a single-mode cavity field in the system. The cavity field is driven by $L$ pulsed driving fields with the amplitudes $\Omega^{(1)}(t)$, $\Omega^{(2)}(t)$, ..., $\Omega^{(L)}(t)$ and the corresponding carrier frequencies $\omega_{1}$, $\omega_{2}$, ..., $\omega_{L}$, as shown in Fig.~\ref{Fig1}(a). In realistic cases, the number of $L$ is determined by the detailed state-generation task. The Hamiltonian of the system reads ($\hbar=1$)
\begin{align}\label{eq1}
	H^{(1)}&=\omega _{c}a^{\dag }a+\omega _{M}b^{\dag }b-g_{0}a^{\dag }a(b^{\dag
	}+b) \notag \\ & \hspace{0.3cm}+ \sum_{l=1}^{L}\Big[\Omega^{(l)}(t) a^{\dag
	}e^{-i\omega _{l}t}+\mathrm{H.c.}\Big],
\end{align}
where $a$ ($a^{\dag}$) and $b$ ($b^{\dag}$) are the annihilation (creation) operators of the cavity field and the mechanical mode, respectively, with the corresponding resonance frequencies $\omega _{c}$ and $\omega _{M}$. The $g_{0}$ is the single-photon optomechanical-coupling strength. In a rotating frame with respect to $\omega _{c}a^{\dag }a$, the Hamiltonian~(\ref{eq1}) becomes
\begin{align}\label{eq2}
	H_{R}^{(1)}\!=\!\omega _{M}b^{\dag }b\!-\!g_{0}a^{\dag }a(b^{\dag
	}+b)\! +\!\sum_{l=1}^{L}\Big[\Omega ^{(l)}(t) a^{\dag
	}e^{-i\Delta _{l}t}+\mathrm{H.c.}\Big],
\end{align}
where $\Delta _{l}=\omega _{l}-\omega _{c}$ for $l=1,2,...,L$, denotes the detunings between the carrier frequencies of the pulsed driving fields and the cavity-field frequency.

We consider the weak driving of the cavity mode, allowing the driving fields to be treated as a perturbation. By using the conditional displacement operator $D_{1}(\beta a^{\dag }a)=\mathrm{exp}[\beta a^{\dag }a(b^{\dag}-b)]$ with $\beta=g_{0}/\omega _{M}$, the first two terms of Eq.~(\ref{eq2}), $H_{S}^{(1)}=\omega _{M}b^{\dag }b-g_{0}a^{\dag }a(b^{\dag}+b)$, can be diagonalized as
\begin{equation}\label{eq3}
	H_{S}^{(1)}=\sum_{m,n=0}^{\infty }E_{m,n}\vert m,\tilde{n}(
	m) \rangle \langle m,\tilde{n}(m) \vert,
\end{equation}
where the eigenstates of $H_{S}^{(1)}$ are $\vert m,\tilde{n}(m) \rangle=\vert m\rangle_{a}\otimes \vert \tilde{n}(m)\rangle_{b}$, with $\vert \tilde{n}(m)\rangle_{b}\equiv D_{1}(m\beta)\vert n\rangle_{b}$. The corresponding eigenvalues are $E_{m,n}=n\omega _{M}-m^{2}\chi$, with $\chi=g_{0}^{2}/\omega _{M}$. Here $\vert m\rangle_{a}$ ($m=0,1,2,...$) represents the Fock states of the cavity mode and $\vert \tilde{n}(m)\rangle_{b}$ denotes the $m$-photon displaced Fock states of the mechanical resonator. In the special case of $m=0$, the eigenstates are reduced to $\vert 0,\tilde{n}(0) \rangle=\vert 0\rangle_{a}\otimes \vert n\rangle_{b}$. The completeness of the eigenstates can be expressed as $\sum_{m,n=0}^{\infty }\vert m,\tilde{n}(m) \rangle\langle m,\tilde{n}(m) \vert=I_{a}\otimes I_{b}$, where $I_{a}$ and $I_{b}$ are the identity operators in the Hilbert space of the cavity field and the mechanical mode, respectively.

In the eigenrepresentation of $H_{S}^{(1)}$, the Hamiltonian $H_{R}^{(1)}$ can be expressed as
\begin{align}\label{eq4}
	H_{R}^{(1)}&=H_{S}^{(1)}+\sum_{l=1}^{L}\sum_{m,n,s=0}^{\infty } \Big[
	A_{n,s}^{(m)}\Omega ^{(l)}( t) e^{-i\Delta _{l}t}  \notag \\ &\hspace{0.3cm} \times \vert m,\tilde{n%
	}( m) \rangle \langle m-1,\tilde{s}( m-1)
	\vert +\mathrm{H.c.}\Big].
\end{align}
Here the coefficients $A_{n,s}^{(m)}$ are defined as $\sqrt{m}_{b}\langle n|D_{1}(-\beta)|s\rangle_{b}$ and can be explicitly calculated as~\cite{OliveiraPRA1990}
\begin{align}\label{eq5}
	A_{n,s}^{(m)}=\begin{cases}
		\sqrt{m}\sqrt{\frac{n!}{s!}}e^{-\beta^{2}/2}\beta^{s-n}L_{n}^{s-n}(\beta^{2}), & n<s\\
		\sqrt{m}\sqrt{\frac{s!}{n!}}e^{-\beta^{2}/2}(-\beta)^{n-s}L_{s}^{n-s}(\beta^{2}), & n\geq s,
	\end{cases}
\end{align}
where $L_{n}^{s}(x)$ represents the associated Laguerre polynomials. In the rotating frame with respect to $H_{S}^{(1)}$, the Hamiltonian~(\ref{eq4}) becomes
\begin{align}\label{eq6}
	H_{I}^{(1)}&=\sum_{l=1}^{L}\sum_{m,n,s=0}^{\infty }\Big[
	A_{n,s}^{(m)}\Omega^{(l)}( t) e^{i(\delta_{m,n,s}-\Delta _{l})t}  \notag \\ &\hspace{0.3cm} \times \vert m,\tilde{n%
	}( m) \rangle \langle m-1,\tilde{s}(m-1)
	\vert +\mathrm{H.c.}\Big],
\end{align}
where we introduce $\delta_{m,n,s}=E_{m,n}-E_{m-1,s}=(n-s)\omega _{M}-(2m-1)\chi$.

\subsection{Effective Hamiltonian within a restricted Hilbert space}
To prepare phononic Fock states and superposed Fock states, we derive the effective Hamiltonian within a finite-dimensional Hilbert space. In the resolved-sideband regime, where the cavity-field decay rate $\kappa$ is smaller than the mechanical frequency $\omega_{M}$, the carrier frequencies $\omega_{l}$ of the pulsed driving fields $\Omega^{(l)}(t)$ are chosen to match the resonance transitions $\vert 0,n\rangle \overset{\Omega^{(1)}(t)}{\longleftrightarrow} \vert 1,\tilde{n}%
(1)\rangle $ and $\vert 1,\tilde{n}(1)\rangle \overset{\Omega^{(l\neq1)}(t)}{\longleftrightarrow}
\vert 0,n+N_{l}\rangle $, with detunings $-\chi$ and $-\chi-N_{l}\omega_{M}$, respectively, as shown in Fig.~\ref{Fig1}(b) (note that only the dominant transitions are labeled in this diagram). Here $N_{l}$ denotes the index of the $N_{l}$th phononic Fock states that resonate with the carrier frequency of the pulsed fields $\Omega^{(l)}(t)$. Under these conditions, the interaction Hamiltonian $H_{I}^{(1)}$ can be divided into two parts
\begin{equation}
	H_{I}^{(1)}=\tilde{H}_{I}^{(1)}+\tilde{H}_{I}^{\prime(1)},
\end{equation}
where $\tilde{H}_{I}^{(1)}$ describes the carrier-frequency resonant-transition part,
\begin{align}\label{eq8}
	\tilde{H}_{I}^{(1)} &=\sum_{n=0}^{\infty }\Big[\Omega ^{(1)}(t)
	A_{n,n}^{(1)}\vert 1,\tilde{n}(1) \rangle
	\langle 0,n\vert   \nonumber \\
	&\hspace{0.3cm}+\sum_{l=2}^{L}\Omega ^{(l)}( t)
	A_{n,n+N_{l}}^{(1)}\vert 1,\tilde{n}( 1) \rangle
	\langle 0,n+N_{l}\vert\Big] +\mathrm{H.c.}
\end{align}
and $\tilde{H}_{I}^{\prime(1)}$ represents the off-resonance-transition part
\small{
\begin{align}\label{eq9}
	\tilde{H}_{I}^{\prime(1)} &\!=\!\sum_{m,n,s=0}^{\infty}\,' \Bigg\{A_{n,s}^{(m)}\Bigg[\Omega
	^{(1)}( t) e^{i\delta _{m,n,s}^{(1)}t}+\sum_{l=2}^{L}\Omega
	^{(l)}( t) e^{i\delta _{m,n,s}^{(l)}t}\Bigg]  \nonumber \\
	&\hspace{0.3cm}\times \vert m,\tilde{n}( m) \rangle \langle
	m-1,\tilde{s}(m-1)\vert +\mathrm{H.c.}\Bigg\}.
\end{align}}\normalsize
The primed summation in Eq.~(\ref{eq9}) excludes those terms of the Hamiltonian (\ref{eq8}), and the off-resonant detunings $\delta _{m,n,s}^{(1,l)}$ in Eq.~(\ref{eq9}) are introduced as
\begin{subequations}
\begin{eqnarray}
	\delta _{m,n,s}^{(1)} &=&(n-s)\omega _{M}-2\chi%
	( m-1), \label{eq10a}  \\
	\delta _{m,n,s}^{(l)} &=&( n-s+N_{l}) \omega _{M}-2\chi(m-1) \label{eq10b},
\end{eqnarray}
\end{subequations}
with $l\geq2$. For these off-resonant contributions, $s\neq n$ in Eq.~(\ref{eq10a}) and $s\neq n+N_{l}$ in Eq.~(\ref{eq10b}) for $m=1$.

To suppress the off-resonant transition term $\tilde{H}_{I}^{\prime(1)}$, the parameter conditions $\vert \delta _{m,n,s}^{(1,l)}\vert \gg \vert
A_{n,s}^{(m)}\vert \vert \Omega ^{(1,l)}\vert _{\max}$ should be satisfied, where $\vert \Omega ^{(1,l)}\vert _{\max}$ represents the maximal amplitude of the pulsed driving fields $\Omega^{(1,l)}$. In particular, it is essential to suppress the transitions from the states $\vert 1,\tilde{n}(1) \rangle$ to $\vert 2,\tilde{n}(2) \rangle$, which requires
\begin{equation}\label{eq11}
	|\delta _{2,n,s}^{(1,l)}|\gg |A_{n,s}^{(2)}||\Omega ^{(1,l)}|_{\max }.
\end{equation}
Hence, when the parameter condition in Eq.~(\ref{eq11}) is satisfied, the Hamiltonian $H_{I}^{(1)}$ can be approximately reduced to $\tilde{H}_{I}^{(1)}$. Though $\tilde{H}_{I}^{(1)}$ operates in an infinite-dimensional Hilbert space, for realistic simulations, we need to truncate the dimension of the phonon-Fock-state space to a finite dimension $N_{m}$ to maintain computational feasibility and accuracy.

Under the truncation, the effective Hamiltonian of the system becomes
\begin{align}\label{eq12}
	H_{\mathrm{eff}}^{(1)} &\!=\!\sum_{n=0}^{N_{m}-1}\Omega ^{(1)}(t)
	A_{n,n}^{(1)}\vert 1,\tilde{n}(1) \rangle
	\langle 0,n\vert +\sum_{l=2}^{L}\sum_{n=0}^{N_{m}-N_{l}}\Omega ^{(l)}( t)   \nonumber \\
	&\hspace{0.4cm} \times
	A_{n,n+N_{l}}^{(1)}\vert 1,\tilde{n}( 1) \rangle
	\langle 0,n+N_{l}\vert +\mathrm{H.c.}.
\end{align}
We point out that this Hamiltonian~(\ref{eq12}) only describes the transitions from the states $\vert 0, n\rangle$ and $\vert 0, n+N_{l}\rangle$ to the state $\vert 1, \tilde{n}( 1) \rangle$ when $n\leq N_{m}$. When the system starts from the state $\vert 0, 0\rangle$ and $g_{0}$ is appropriately chosen to ensure that $A_{N_{l},N_{l}}^{(1)}\approx0$, the pulse $\Omega ^{(1)}(t)$ mainly drives the transition from $\vert 0, 0\rangle$ to $\vert 1,\tilde{0}(1) \rangle$. Meanwhile, the transition from the state $\vert 0, N_{l}\rangle$ to $\vert 1,\tilde{N_{l}}( 1) \rangle$ is negligible due to the suppressed transition-matrix element $\Omega ^{(1)}(t)A_{N_{l},N_{l}}^{(1)}$. The pulses $\Omega ^{(2)}(t)$, $\Omega ^{(3)}(t)$,..., $\Omega ^{(L)}(t)$ are applied simultaneously, with each pulse preselected with appropriate resonance conditions to drive transitions from the state $\vert 1,\tilde{0}(1) \rangle$ to the desired states. Specifically, if the goal is to prepare an $N_{2}$ phononic Fock state, only one additional pulse ($L=2$) is needed, which drives the transition from the state $\vert 1,\tilde{0}(1) \rangle$ to $\vert 0, N_{2}\rangle$. However, if a superposed Fock state $(\vert N_{2}\rangle+\vert N_{3}\rangle)/\sqrt{2}$ is desired, two additional pulses are required: One pulse pumps the transition from the state $\vert 1,\tilde{0}(1) \rangle$ to $\vert 0, N_{2}\rangle$ and the other drives the transition from the state $\vert 1,\tilde{0}(1) \rangle$ to $\vert 0, N_{3}\rangle$. Note that although the dissipation can induce transitions to other energy levels, the populations of these levels are relatively small, and therefore the aforementioned transitions remain dominant.
\begin{figure}[tbp]
	\centering \includegraphics[width=0.48\textwidth]{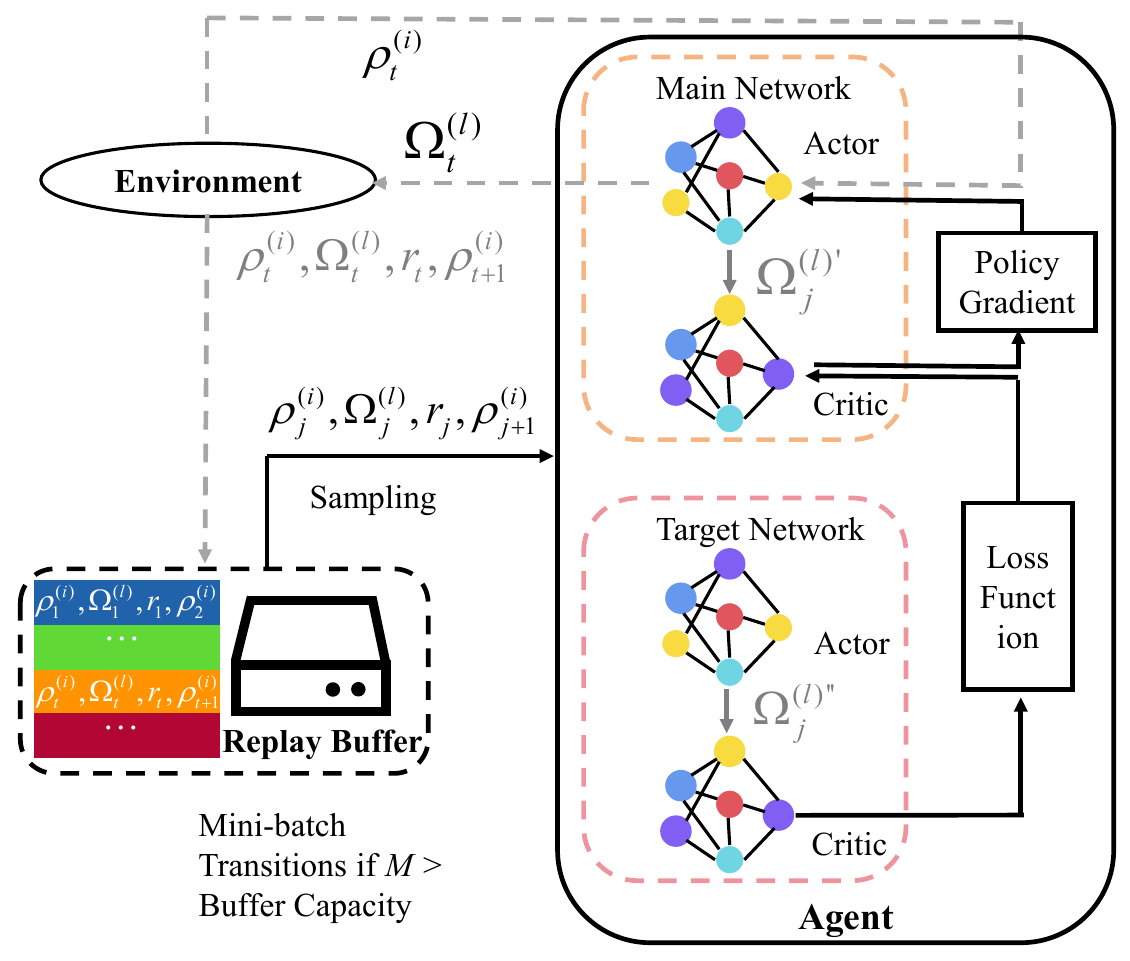}
	\caption{Flowchart of the deep deterministic policy gradient algorithm applied in this work. The agent consists of both the main and target networks to optimize the control strategy. The main network generates control actions $\Omega_{t}^{(l)}$ based on the input state $\rho^{(i)}_{t}$, while the critic evaluates these actions. The target network helps stabilize the learning process. The environment, representing the driven-free optomechanical system, evolves the quantum state $\rho^{(i)}_{t}$ to $\rho^{(i)}_{t+1}$, and the reward $r_{t}$ is calculated based on the fidelity. State-action-reward tuples $(\rho^{(i)}_{t}, \Omega^{(l)}_{t}, r_{t}, \rho^{(i)}_{t+1})$ are stored in the replay buffer. After the $M$th epoch, a random sample $(\rho^{(i)}_{j}, \Omega^{(l)}_{j}, r_{j}, \rho^{(i)}_{j+1})$ is drawn, where $j$ indexes the sampled data used to update both networks.}
	\label{Fig2}
\end{figure}

\subsection{Dynamical evolution of the optomechanical system in the open-system case}
\begin{figure*}[tbp]
	\centering \includegraphics[width=1\textwidth]{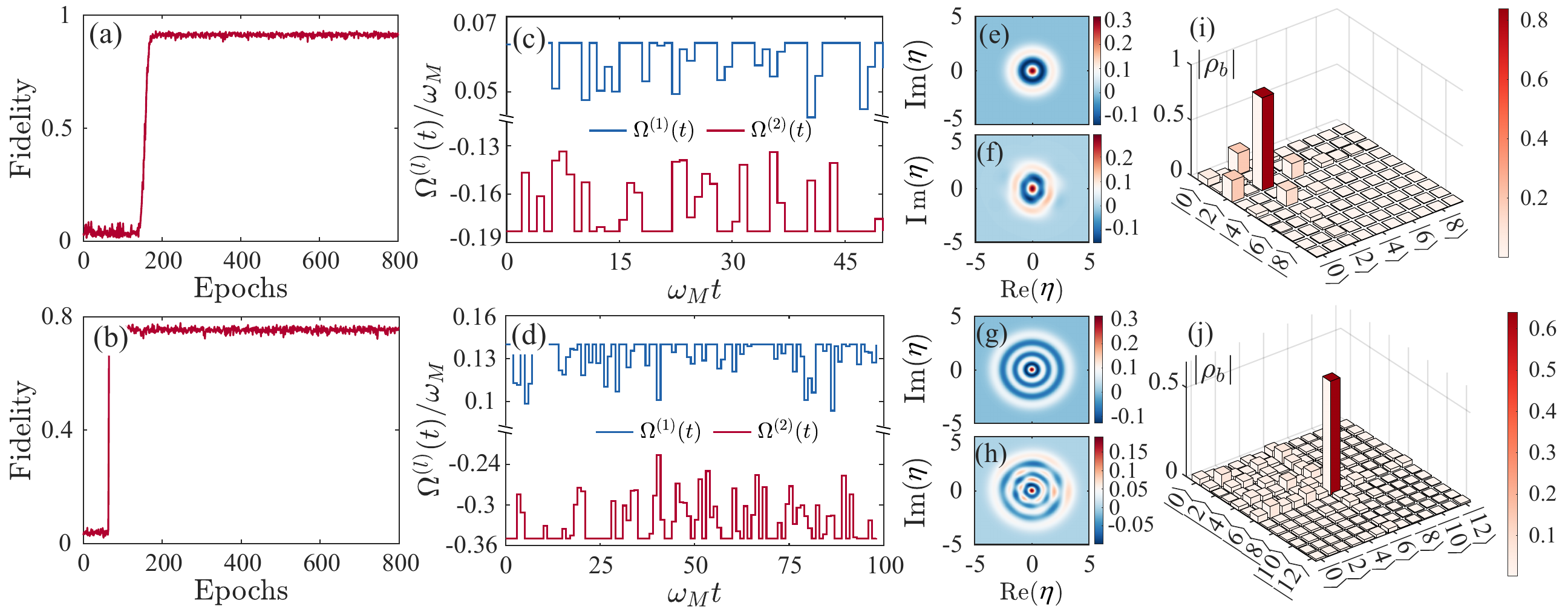}
	\caption{Numerical results for the preparation of the phononic Fock states. (a) and (b) Fidelity as a function of the training epochs for the target states $|N\rangle_{b}$ when (a) $N=2$ and (b) $N=6$. (c) and (d) Scaled driving amplitudes $\Omega^{(1)}(t)/\omega_{M}$ and $\Omega^{(2)}(t)/\omega_{M}$ versus $\omega_{M}t$, corresponding to the maximum fidelity shown in (a) and (b). (e) and (g) Wigner functions of the Fock states $|2\rangle$ and $|6\rangle$, respectively. (f) and (h) Wigner functions of the corresponding simulated Fock states. (i) and (j) Absolute values of the density-matrix elements in the Fock-state representation of the simulated states. Here we take $N_{c}=3$,  $n_{\mathrm{th}}=0$, $\kappa/\omega_{M}=0.002$, and $\gamma_{M}/\omega_{M}=0.0004$ in all panels. The additional parameters are $N_{m}=10$, $g_{0}/\omega_{M}=0.839$, $S=50$, and $\omega_{M}T=50$ in the top panels and $N_{m}=13$, $g_{0}/\omega_{M}=1.752$, $S=98$, and $\omega_{M}T=98$ in the bottom panels.}
	\label{Fig3}
\end{figure*}
In realistic situations, the dissipations will inevitably affect the dynamics of the system. In the single-photon ultrastrong-coupling regime~\cite{LiaoPRA2020,ZouPRA2022}, the ordinary quantum master equation~\cite{scully1997quantum} fails to accurately describe the behavior of the system and needs to be replaced by the dressed master equation~\cite{SettineriPRA2018}. For the optomechanical system governed by $H_{S}^{(1)}= \omega _{M}b^{\dag }b-g_{0}a^{\dag }a(b^{\dag}+b)$, the dissipation of the mechanical mode can be described in the dressed-state representation. In this representation, the transformed operator is given by $e^{iH_{S}^{(1)}t}be^{-iH_{S}^{(1)}t}=e^{-i\omega _{M}t}(b-\beta a^{\dagger}a)+\beta a^{\dagger}a$, where the terms $b-\beta a^{\dagger}a$ and $\beta a^{\dagger}a$ oscillate at well-defined frequencies. However, for the cavity mode, the transformed operator takes the form $e^{iH_{S}^{(1)}t}ae^{-iH_{S}^{(1)}t}=e^{i\beta^{2}[\omega _{M}t-\sin(\omega _{M}t)](2a^{\dagger}a+1)}ae^{\beta[(e^{i\omega _{M}t}-1)b^{\dagger}-(e^{-i\omega _{M}t}-1)b]}$.~This operator involves multiple frequency components, complicating the construction of the Lindblad operator. To address this complexity, the zeroth-order approximation is applied: $ e^{i\beta^{2}[\omega _{M}t-\sin(\omega _{M}t)](2a^{\dagger}a+1)}ae^{\beta[(e^{i\omega _{M}t}-1)b^{\dagger}-(e^{-i\omega _{M}t}-1)b]}\approx a$. This approximation simplifies the oscillation frequency of the cavity mode to $\omega_{c}$, thereby enabling the derivation of the Lindblad operator for the cavity mode. The validity of this approximation is ensured by the large frequency separation between the cavity field and the mechanical oscillation, i.e., $\omega_{c}$ ($\sim 10^{14}$~$\mathrm{Hz}$) is much larger than $\omega_{M}$ ($\sim 10^{7}$~$\mathrm{Hz}$). In the derivation of the dressed master equation, we assume an Ohmic spectral density for the mechanical bath and treat the pulse driving terms as perturbations.

Based on these approximations, the dressed master equation for the driven optomechanical system in the single-photon ultrastrong-coupling regime is obtained as~\cite{HuPRA2015}
\begin{align}\label{eq13}
\dot{\rho}^{(1)}(t)&=i[\rho^{(1)}(t),H_{R}^{(1)}]+\gamma_{M}(n_{\text{th}}+1)\mathcal{D}[b-\beta a^{\dagger}a]\rho^{(1)}(t)\nonumber\\
&\quad+\gamma_{M}n_{\text{th}}\mathcal{D}[b^{\dagger}-\beta a^{\dagger}a]\rho^{(1)}(t)+\kappa\mathcal{D}[a]\rho^{(1)}(t)\nonumber\\
&\quad+4\gamma_{M}(k_{B}T_{b}/\omega_{M})\beta^{2}\mathcal{D}[a^{\dagger}a]\rho^{(1)}(t),
\end{align}
where $\rho^{(1)}(t)$ is the density matrix of the single-resonator optomechanical system, $H_{R}^{(1)}$ is defined in Eq.~(\ref{eq2}), and $\mathcal{D}[o]\rho^{(1)} (t)$ is the Lindblad superoperator given by
$\mathcal{D}[o]\rho^{(1)} (t)=[2o\rho^{(1)} (t)o^{\dag }-\rho^{(1)} (t)o^{\dag }o-o^{\dag
}o\rho^{(1)} (t)]/2$. The parameters $\kappa$ and $\gamma_{M}$ denote the decay rates of the cavity field and the mechanical mode, respectively. The thermal phonon occupation number at temperature $T_{b}$ is given by $n_{\text{th}}=[\text{exp}(\hbar\omega_{M}/k_{B}T_{b})-1]^{-1}$, where $k_{B}$ is the Boltzmann constant. By numerically solving Eq.~(\ref{eq13})~\cite{JOHANSSON20131234}, the transient density operator $\rho^{(1)} (t)$ of the system can be obtained.

\subsection{Generation of phononic Fock states}
In this section, we exhibit how to generate the phononic Fock states with the above-introduced method. Concretely, we present the results for $N=2$ and $6$ as examples. To generate the two- and six-phonon Fock states in the system, we need two pulsed driving fields ($L=2$). The resonant detunings of the pulsed driving fields $\Omega^{(2)}(t)$ for the cases of $N=2$ and $6$ are taken as $-\chi-2\omega_{M}$ and $-\chi-6\omega_{M}$, respectively. Due to the complex dynamics of the transitions induced by both the pulsed drivings and dissipations, it is a difficult mission to design the optimized driving parameters for pursuing a high-fidelity state generation. Here we adopt the reinforcement learning method to realize the optimization of the driving fields.

To implement the above-mentioned reinforcement learning, a main neural network is introduced to determine the optimal control actions for the environment, aiming to maximize a predefined reward function. We point out that here the term “environment" for reinforcement learning refers to the undriven single-resonator optomechanical system, with its dynamical evolution governed by Eq.~(\ref{eq13}), rather than the real bath of the physical system. The workflow for our system is displayed in Fig.~\ref{Fig2}. At each time step $t=s\Delta t$ ($s = 1, 2,..., S$), where the total evolution time $T$ is uniformly divided into $S$ intervals of duration $\Delta t=T/S$, the environment provides the current quantum state $\rho^{(i)}_{t}$ to the agent. Here the superscript $i=1,2$ in the density operator $\rho^{(i)}_{t}$ distinguishes between the single-resonator ($i=1$) and two-resonator ($i=2$) optomechanical systems. The agent then generates control actions $\Omega^{(l)}_{t}$ for the $l$th pulsed field [multiple pulsed fields can be generated as required, with each one denoted by the superscript $(l)$]. These control amplitudes are then applied back to the environment, driving it to evolve into a new quantum state $\rho^{(i)}_{t+1}$, which is fed back to the agent, completing one interaction cycle (a full loop of state, action, and feedback). This iterative process allows the agent to refine its control strategy over time.

\begin{figure}[tbp]
	\centering \includegraphics[width=0.48\textwidth]{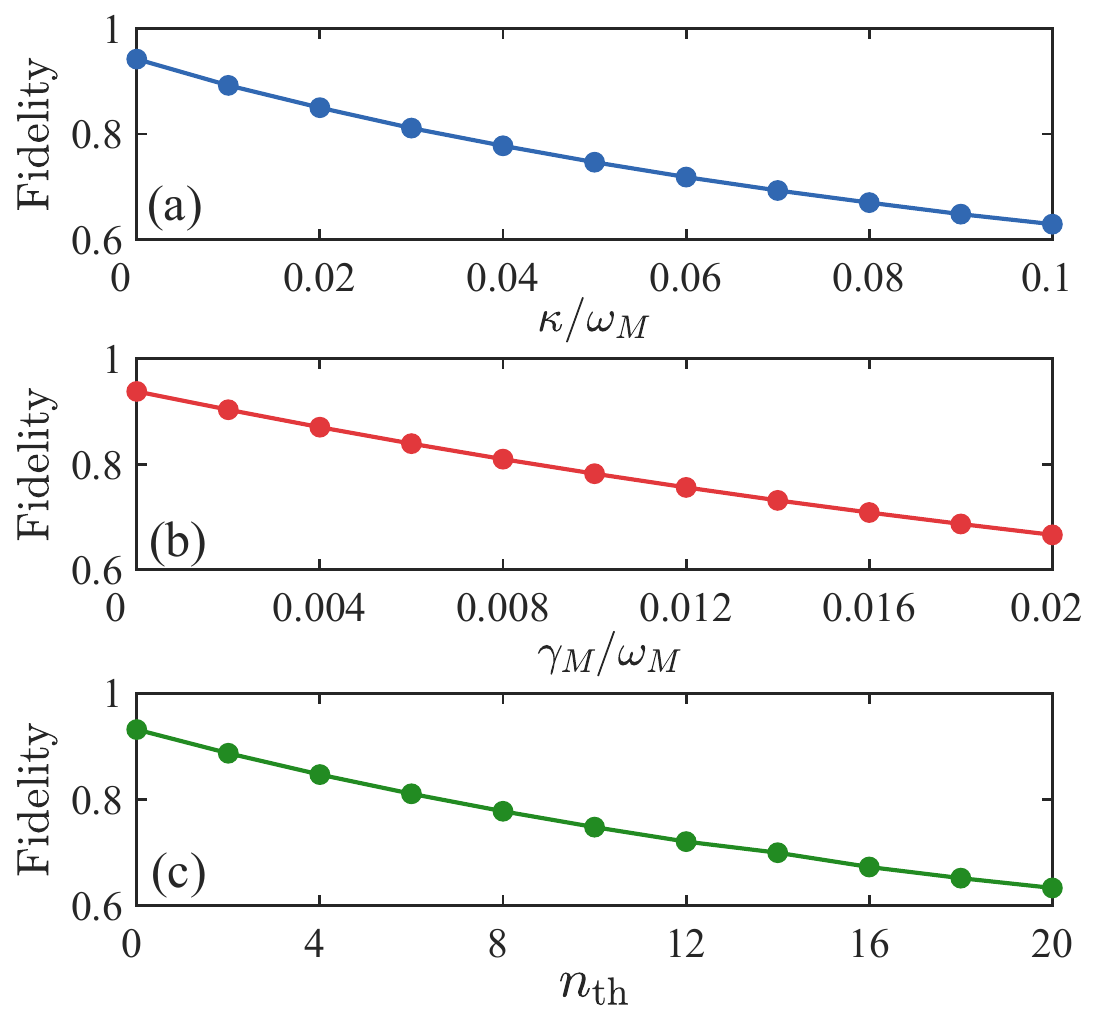}
	\caption{Maximum fidelity of the phononic Fock state $|2\rangle_{b}$ after 800 iterations as a function of (a) scaled cavity-field decay rate $\kappa/\omega_{M}$, (b) scaled mechanical decay rate $\gamma_{M}/\omega_{M}$, and (c) thermal phonon occupation $n_{\mathrm{th}}$ of the mechanical resonator. The parameters are $N_{c}=3$, $N_{m}=10$, $g_{0}/\omega_{M}=0.839$, $\omega_{M}T=50$, and (a) $n_{\mathrm{th}}=0$ and $\gamma_{M}/\omega_{M}=0.0004$, (b) $n_{\mathrm{th}}=0$ and $\kappa/\omega_{M}=0.002$, and (c) $\kappa/\omega_{M}=0.002$, $\gamma_{M}/\omega_{M}=0.0004$.}
	\label{Dispa}
\end{figure}

The DDPG algorithm is utilized to train the agent. It is based on the actor-critic framework and incorporates several key components, including a reply buffer and separate target network for both the actor and critic (Fig.~\ref{Fig2}). During the initial $M$ epochs, the pulsed fields $\Omega^{(l)}_{t}$ are generated randomly by the main actor network. The resulting state-action-reward tuples $(\rho^{(i)}_{t}, \Omega^{(l)}_{t}, r_{t}, \rho^{(i)}_{t+1})$ are stored in the replay buffer for later use. Here, $\rho^{(i)}_{t}$ and $\rho^{(i)}_{t+1}$ represent the quantum states at consecutive time steps $t$ and $t+1$, $\Omega^{(l)}_{t}$ represents the control amplitude of the $l$th pulsed field, and $r_{t}=-10\log_{10}(1-F)$ is the reward, where $F=\langle \psi|\rho^{(i)}_{t+1}|\psi\rangle$ is the fidelity between the simulated state $\rho^{(i)}_{t+1}$ and the target state $|\psi\rangle$. After the initial exploration stage, the replay buffer is used to randomly select batches of 128 samples for training. These samples are fed into both the main and target networks to calculate the loss and update the model parameters. The target networks are updated at a slower rate by absorbing a fraction (e.g., 10\%) of the weights from the main networks, ensuring stable and gradual learning.

To enable neural network processing, each quantum state $\rho^{(i)}_{t}$ is transformed into a suitable input format. Specifically, $\rho^{(i)}_{t}$ is a complex-valued matrix of dimension $D\times D$, where $D=N_{c}\times N_{m}$ denotes the total Hilbert space dimension, with $N_{c}$ and $N_{m}$ the dimensions of the cavity and mechanical modes, respectively. This matrix is vectorized by separating the real and imaginary components, resulting in a real-valued vector of length  $2D^{2}$. The training process is conducted over $E = 800$ epochs, each consisting of $S$ time steps. Since each time step involves solving the quantum master equation, the total training time depends on $D$, $S$, and $E$ and typically takes several hours. For reproducibility, we fix the random seed and periodically save the model parameters during training.

\begin{figure*}[tbp]
	\centering \includegraphics[width=1\textwidth]{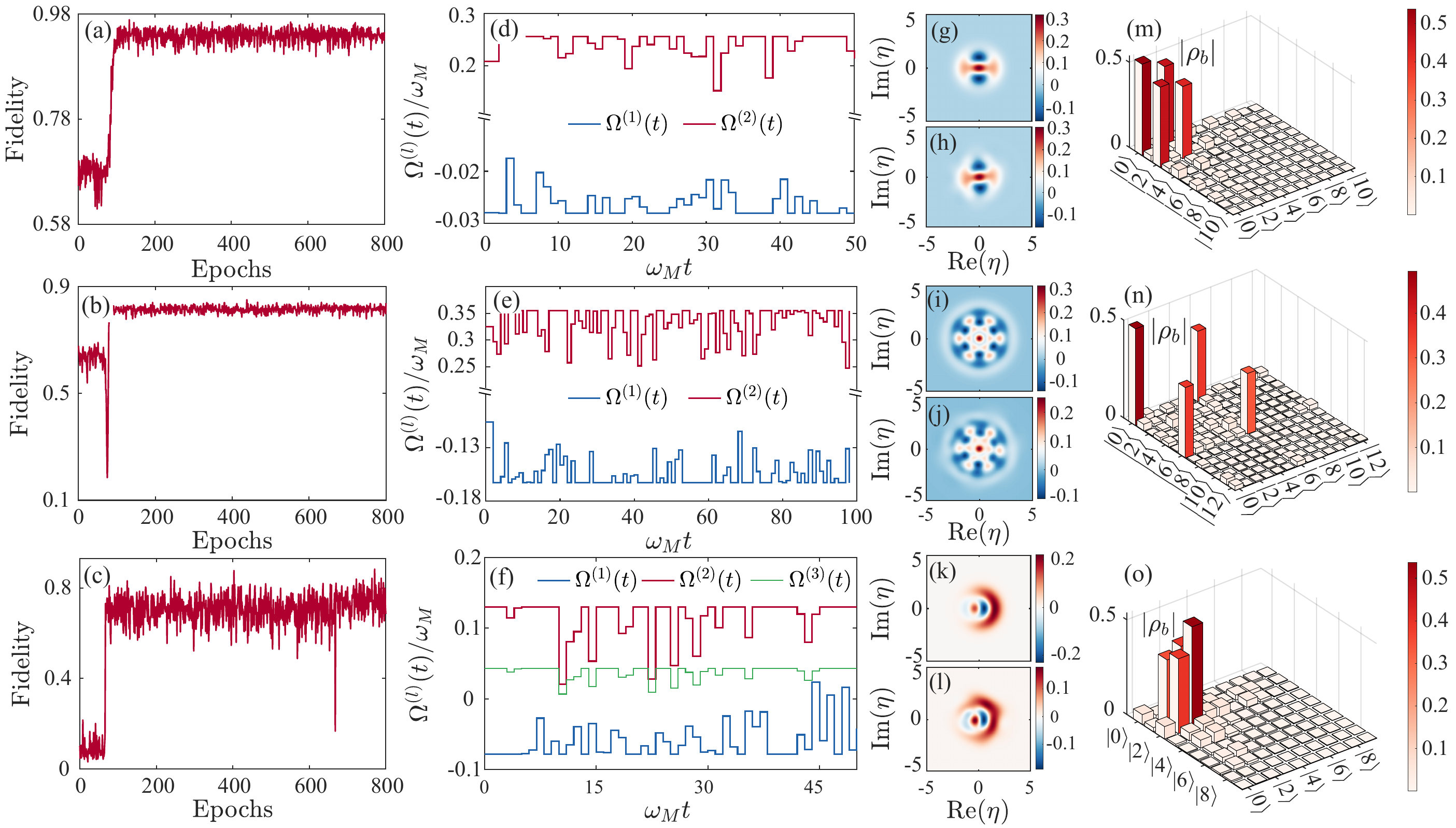}
	\caption{Numerical results for the generation of the superposed Fock states. Fidelity is plotted as a function of the training epochs for the target states (a) $(|0\rangle_{b}+|2\rangle_{b})/\sqrt{2}$, (b) $(|0\rangle_{b}+|6\rangle_{b})/\sqrt{2}$, and (c) $(|1\rangle_{b}+|2\rangle_{b})/\sqrt{2}$. (d)-(f) Scaled driving amplitudes $\Omega^{(l=1,2,3)}(t)/\omega_{M}$ plotted as a function of $\omega_{M}t$, corresponding to the maximum fidelity shown in (a)-(c). (g), (i), and (k) Wigner functions of  ideal and (h), (j), and (l) simulated superposed Fock states. (m)-(o) Absolute value of the density-matrix elements in the Fock-state representation for the evolved mechanical states. Here we take $N_{c}=3$, $n_{\mathrm{th}}=0$, $\kappa/\omega_{M}=0.002$, and $\gamma_{M}/\omega_{M}=0.0004$ in all panels. The other parameters are $N_{m}=11$, $g_{0}/\omega_{M}=0.78$, $S=50$, and $\omega_{M}T=50$ in the top panels; $N_{m}=13$, $g_{0}/\omega_{M}=1.716$, $S=98$, and $\omega_{M}T=98$ in the middle panels; and $N_{m}=10$, $g_{0}/\omega_{M}=0.89$, $S=50$, and $\omega_{M}T=50$ in the bottom panels.}
	\label{Fig4}
\end{figure*}
To exhibit the performance of the state generation, we present the results for the generation of the Fock states $|2\rangle_{b}$ and $|6\rangle_{b}$ in the top and bottom rows of Fig.~\ref{Fig3}, respectively. Figures~\ref{Fig3}(a) and \ref{Fig3}(b) show the fidelity as a function of the training epochs. For $N=2$, the maximum fidelity of $0.935$ is reached after $610$ epochs, while for $N=6$, the maximum fidelity of $0.778$ is achieved after $247$ epochs. This is because for a larger $N$, it becomes very difficult to find the optimized fields corresponding to a high fidelity due to the larger dimension of the Hilbert space involved and the very complex state transitions. The optimized pulsed driving amplitudes, $\Omega^{(1)}(t)/\omega_{M}$ and $\Omega^{(2)}(t)/\omega_{M}$  corresponding to the epoch with maximum fidelity in Figs.~\ref{Fig3}(a) and~\ref{Fig3}(b) are shown in Figs.~\ref{Fig3}(c) and~\ref{Fig3}(d), respectively. The peak values satisfy the condition~(\ref{eq11}), ensuring that the Hilbert space of the cavity mode could be truncated to a low-dimensional subspace.

The nonclassicality of the generated phononic Fock states can be examined by inspecting the Wigner functions $\mathrm{W}(\eta)=\frac{2}{\pi}\mathrm{Tr}[D_{b}^{\dagger}(\eta)\rho_{b}D_{b}(\eta)(-1)^{b^{\dagger}b}]$~\cite{scully1997quantum},displayed in Figs.~\ref{Fig3}(f) and~\ref{Fig3}(h) for $N=2$ and $6$, respectively. Here $D_{b}(\eta)=\text{exp}(\eta b^{\dagger}-\eta^{\ast}b)$ is the displacement operator and the density matrix $\rho_{b}$ is obtained by taking the trace of the system density operator over the cavity mode, specifically as $\rho_{b}=\mathrm{Tr}_{a}(\rho^{(1)})$. By comparing the simulated Wigner functions with those for the ideal Fock states [Figs.~\ref{Fig3}(e) and~\ref{Fig3}(g)], we find that the plotted Wigner functions can exhibit the main feature of the Fock states. The minor deviation is caused by the mixing of other Fock states due to the existence of other transitions. To further validate the results, the absolute values of the density-matrix elements $\rho_{b}$ in the Fock-state representation are shown in Figs.~\ref{Fig3}(i) and~\ref{Fig3}(j). Here the diagonal elements represent the populations of different phononic Fock states. The results confirm that the target state has the highest population. Meanwhile, we can also see some slight mixing of other adjacent Fock states. The negligible populations in the larger Fock states can validate the low-dimensional truncation of the mechanical mode and confirm the accuracy of the state preparation.

In our previous simulations, the cavity-field decay rate $\kappa$, the mechanical decay rate $\gamma_{M}$, and the mechanical thermal phonon occupation $n_{\mathrm{th}}$ were fixed. However, these parameters are tunable and significantly influence the preparation of nonclassical states. In Fig.~\ref{Dispa}, we show the maximum fidelity of the phononic Fock state $|2\rangle_{b}$ after 800 iterations as a function of these parameters. As shown, increasing $\kappa$ [Fig.~\ref{Dispa}(a)], $\gamma_{M}$ [Fig.~\ref{Dispa}(b)], or $n_{\mathrm{th}}$ [Fig.~\ref{Dispa}(c)] results in a decrease in the maximally achievable fidelity. The reduction in fidelity caused by an increase in $\kappa$ is due to enhanced photon loss, which increases the transition probability from the single-photon state to the zero-photon state. This leads to phonon state mixing during the dynamics, thereby decreasing the fidelity of the target state. On the other hand, increases in $\gamma_{M}$ and $n_{\mathrm{th}}$ directly reduce the population of the target phononic Fock state, further lowering the fidelity. Additionally, for the superposed Fock states and two-mode entanglement states discussed later, increasing $\kappa$, $\gamma_{M}$, and $n_{\mathrm{th}}$ similarly reduces the fidelity of the prepared states. Therefore, we do not further discuss the effects of these parameters on the states presented later.

\subsection{Generation of superposed Fock states}
In this section, we investigate the generation of superposed Fock states. We carefully select resonant transition frequencies and apply reinforcement learning to optimize the pulsed driving fields, thereby suppressing unwanted transitions and achieving high fidelity. For the superposed Fock states $(|0\rangle_{b}+|2\rangle_{b})/\sqrt{2}$ and $(|0\rangle_{b}+|6\rangle_{b})/\sqrt{2}$, we keep the same number of pulsed driving fields as in the Fock-state protocols for $|2\rangle_{b}$ and $|6\rangle_{b}$. The pulse shapes are re-optimized via reinforcement learning to ensure that part of the population remains in $|0\rangle_{b}$, creating a coherent superposition rather than transferring all population to $|N\rangle_{b}$. In contrast, for the superposed Fock state $(|1\rangle_{b}+|2\rangle_{b})/\sqrt{2}$, we introduce three pulsed driving fields with the carrier-wave resonant detunings of $-\chi$, $-\chi-\omega_{M}$, and $-\chi-2\omega_{M}$. Here, the pulsed field with resonant detuning $-\chi$ drives the transition from $\vert 0, 0\rangle$ to $\vert 1,\tilde{0}(1) \rangle$, while the additional fields with detunings $-\chi-\omega_{M}$ and $-\chi-2\omega_{M}$ drive the transitions from $\vert 1,\tilde{0}(1) \rangle$ to $\vert 0, 1\rangle$ and from $\vert 1,\tilde{0}(1) \rangle$ to $\vert 0, 2\rangle$, respectively. The coordinated action of these three fields ensures that the population is appropriately distributed between $\vert 0, 1\rangle$ and $\vert 0, 2\rangle$, forming the desired coherent superposition.

To illustrate the results for the generation of three target superposed Fock states: $(|0\rangle_{b}+|2\rangle_{b})/\sqrt{2}$ (top row), $(|0\rangle_{b}+|6\rangle_{b})/\sqrt{2}$ (middle row), and $(|1\rangle_{b}+|2\rangle_{b})/\sqrt{2}$ (bottom row), we plot the fidelities against the training epochs in Figs.~\ref{Fig4}(a)-\ref{Fig4}(c). Here we can see that the maximum fidelities for these three states are 0.955, 0.851, and 0.883, respectively, demonstrating the validity of reinforcement learning in optimizing the state-preparation process. Similar to the Fock-state generation, the fidelity is lower when a larger Fock-state component is involved in the superposed Fock states. It is important to note that DDPG training includes exploration of noise in the actor policy, which can cause temporary drops in fidelity when new driving‑pulse parameters are sampled. However, the fidelity recovers quickly as the neural networks update and converge towards a near‑optimal solution. The optimized driving fields for each state are shown in Figs.~\ref{Fig4}(d)-\ref{Fig4}(f). Here we can see that these fields have complex time-dependent structures, which indicates that the precise control of the fields is required to generate the target states. Importantly, the driving amplitudes satisfy the condition~(\ref{eq11}), ensuring that the cavity mode remains within a low-dimensional subspace.

To verify the quantum coherence of the generated superposed Fock states, the Wigner functions of the simulated mechanical states are shown in Figs.~\ref{Fig4}(h),~\ref{Fig4}(j), and~\ref{Fig4}(l). These Wigner functions exhibit characteristic interference fringes, indicating a signature of quantum superposition. By comparing these Wigner functions with those for the target states [Figs.~\ref{Fig4}(g),~\ref{Fig4}(i), and~\ref{Fig4}(k)], we find that these Wigner functions are closely aligned with those of the target states. Minor deviations are observed due to small populations in these unwanted phonon states. The absolute value of the density-matrix elements in the Fock-state representation are presented in Figs.~\ref{Fig4}(m)-\ref{Fig4}(o). The main diagonal elements indicate that the target components have the highest populations, confirming the successful preparation of the superposed Fock states. For example, in the state $(|0\rangle_{b}+|2\rangle_{b})/\sqrt{2}$, both of the components $|0\rangle_{b}$ and $|2\rangle_{b}$ take approximately half of the total population. Negligible populations in higher phonon states validate the effectiveness of the mechanical-mode truncation. Additionally, the off-diagonal elements highlight the coherence between the target superposition components, further demonstrating the quantum superposition nature of the states and distinguishing them from the statistical mixtures of these superposition components.

\section{Generation of two-mode entangled states in the two-mechanical-resonator optomechanical system\label{Sec3}}
In this section, we generalize the above-introduced state-generation method to create the two-mode entangled states involving two mechanical modes in a two-mechanical-resonator optomechanical system. In the following, we introduce the physical model and Hamiltonians. We also derive the effective Hamiltonian in a confined Hilbert space. Finally, we employ the reinforcement learning method to optimize the pulsed driving fields for high-fidelity preparation of the two-mode entangled state in the open-system case.

\subsection{Physical model and Hamiltonians}
\begin{figure}[tbp]
	\centering \includegraphics[width=0.5\textwidth]{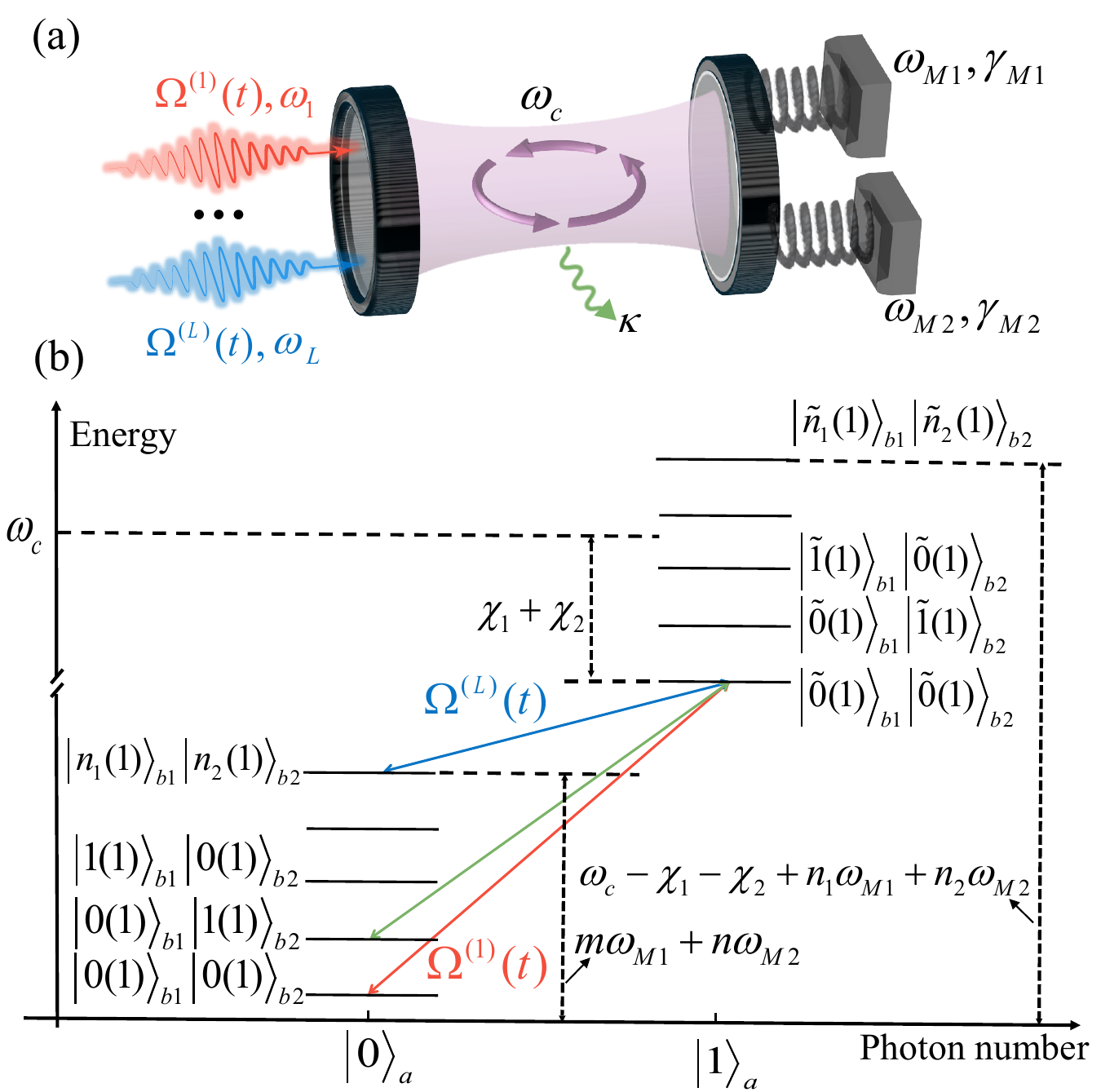}
	\caption{(a) Schematic diagram of a two-resonator optomechanical system. The two mechanical resonators have resonance frequencies $\omega_{M1}$ and $\omega_{M2}$ and corresponding decay rates $\gamma_{M1}$ and $\gamma_{M2}$. Other system parameters are defined in Fig.~\ref{Fig1}. (b) Unscaled energy-level structure of the two-resonator optomechanical system restricted in the zero- and one-photon subspaces. These transitions among the effective energy levels are caused by these driving pulses.}
	\label{Fig5}
\end{figure}

We consider a three-mode optomechanical system consisting of a cavity field optomechanically coupled to two mechanical resonators [Fig.~\ref{Fig5}(a)]. The Hamiltonian of the system takes the form 
\begin{align}\label{eq14}
	H^{(2)}&=\omega _{c}a^{\dag }a+\sum_{i=1,2}\omega _{Mi}b^{\dag }_{i}b_{i}-\sum_{i=1,2}g_{0i}a^{\dag }a(b^{\dag
	}_{i}+b_{i}) \notag \\ & \hspace{0.3cm}+ \sum_{l=1}^{L}\Big[\Omega^{(l)}(t) a^{\dag
	}e^{-i\omega _{l}t}+\mathrm{H.c.}\Big],
\end{align}
where $b_{i=1,2}$ ($b_{i=1,2}^{\dagger}$) is the annihilation (creation) operators for the $i$th mechanical mode, with the resonance frequencies $\omega _{Mi}$ and single-photon optomechanical-coupling strength $g_{0i}$. Other parameters were introduced in Eq.~(\ref{eq1}). Similarly, we turn to a rotating frame with respect to $\omega_{c}a^{\dag }a$; then the Hamiltonian becomes
\begin{align}\label{eq15}
	H_{R}^{(2)}&=\sum_{i=1,2}\omega _{Mi}b^{\dag }_{i}b_{i}-\sum_{i=1,2}g_{0i}a^{\dag }a(b^{\dag
	}_{i}+b_{i}) \notag \\ & \hspace{0.3cm}+
    \sum_{l=1}^{L}\Big[\Omega^{(l)}(t) a^{\dag
	}e^{-i\Delta_{l}t}+\mathrm{H.c.}\Big].
\end{align}
These driving parameters were introduced in Eq.~(\ref{eq2}).

To diagonalize the optomechanical-coupling terms, we introduce the displacement operators $D_{2}(\beta _{1}a^{\dag }a,\beta _{2}a^{\dag }a)=\prod_{i=1}^{2}\mathrm{exp}[\beta _{i}a^{\dag }a(b_{i}^{\dag }-b_{i})]$, where $\beta _{i}=g_{0i}/\omega_{Mi}$ represent the displacement of the $i$th resonator induced by a single photon. By applying this transformation, the first two terms of Eq.~(\ref{eq15}) can be diagonalized as
\begin{equation}\label{eq16}
	H_{S}^{(2)}\!=\!\sum_{i=1,2}\omega _{Mi}b^{\dag }_{i}b_{i}-\sum_{i=1,2}\chi_{i} (a^{\dag }a)^{2},
\end{equation}	
where $\chi_{i}=g_{0i}^{2}/\omega_{Mi}$ for $i=1,2$ are the nonlinear energy shifts caused by the optomechanical couplings. Expanding Hamiltonian~(\ref{eq16}) in terms of its eigenstates, we obtain
\begin{equation}\label{eq17}
	\small
	H_{S}^{(2)}\!=\!\sum_{m,n_{1},n_{2}=0}^{\infty }E_{m,n_{1},n_{2}}\vert m, \tilde{n}_{1}(m),\tilde{n}_{2}(m)\rangle \langle m, \tilde{n}_{1}(m),\tilde{n}_{2}(m) \vert,
\end{equation}%
where $\vert m, \tilde{n}_{1}(m),\tilde{n}_{2}(m)\rangle \equiv\vert m\rangle _{a}\otimes\vert \tilde{n}_{1}(m) \rangle _{b1}\otimes\vert \tilde{n}_{2}( m)\rangle _{b2}$ are the eigenstates of the Hamiltonian $H_{S}^{(2)}$, with $\vert \tilde{n}_{1}(m) \rangle_{b1}=\mathrm{exp}[m\beta_{1}( b_{1}^{\dag }-b_{1})] \vert n_{1}\rangle _{b1}$ and $\vert \tilde{n}_{2}(m)
\rangle _{b2}=\mathrm{exp}[m\beta _{2}( b_{2}^{\dag }-b_{2})]
\vert n_{2}\rangle _{b2}$. The corresponding eigenvalues are given by
\begin{equation}\label{eq18}
	E_{m,n_{1},n_{2}}=n_{1}\omega _{M1}+n_{2}\omega _{M2}-(\chi _{1}+\chi
	_{2}) m^{2}.
\end{equation}
In the presence of the drivings, the transitions will take place among these discrete energy levels, as illustrated in Fig.~\ref{Fig5}(b). Here the eigenstates $\vert m, \tilde{n}_{1}(m),\tilde{n}_{2}(m)\rangle$ also satisfy the completeness relation.

The Hamiltonian $H_{R}^{(2)}$ can be further expressed in the eigenstate representation defined in Eq.~(\ref{eq17}) as
\small{
\begin{align}\label{eq19}	
	H_{R}^{(2)}&\!=\sum_{m,n_{1},n_{2}=0}^{\infty }E_{m,n_{1},n_{2}}\vert m, \tilde{n}_{1}(m),\tilde{n}_{2}(m)\rangle \langle m, \tilde{n}_{1}(m),\tilde{n}_{2}(m) \vert \notag \\
	&\hspace{0.4cm}+\!\sum_{l=1}^{L}\!\sum_{\substack{m,n_{1},n_{2}\\n_{1}^{\prime},n_{2}^{\prime}=0}}^{\infty }\!\Big[\!A_{n_{1},n_{2},n_{1}^{\prime },n_{2}^{\prime}}^{(m)
	}\!\Omega ^{(l)}(t)e^{-i\Delta _{l}t} \vert \!m,\!\tilde{n}_{1}(m) ,\!\tilde{n}_{2}(m)
	\rangle  \notag \\
	&\hspace{0.4cm}  \times  \langle m-1,\tilde{n}_{1}^{\prime }(m-1) ,\tilde{n%
	}_{2}^{\prime }(m-1) \vert+\mathrm{H.c.}\Big].
\end{align}}\normalsize
Here, the coefficients $A_{n_{1},n_{2},n_{1}^{\prime },n_{2}^{\prime}}^{(m)}$ are defined as $A_{n_{1},n_{2},n_{1}^{\prime },n_{2}^{\prime }}^{( m) }=\sqrt{m}_{b1} \langle n_{1}\vert D_{1}( -\beta _{1}) \vert n_{1}^{\prime}\rangle _{b1b2}\langle n_{2}\vert D_{1}( -\beta_{2}) \vert n_{2}^{\prime }\rangle _{b2}$ and the calculation of $A_{n_{1},n_{2},n_{1}^{\prime },n_{2}^{\prime}}^{(m)}$ follows the same procedure as Eq.~(\ref{eq5}). Transforming $H_{R}^{(2)}$ into a rotating frame with respective to $H_{S}^{(2)}$ yields
\small{
	\begin{align}\label{eq20}
	H_{I}^{(2)}&=\sum_{l=1}^{L}\sum_{\substack{m,n_{1},n_{2}\\n_{1}^{\prime},n_{2}^{\prime}=0}}^{\infty }\Big[A_{n_{1},n_{2},n_{1}^{\prime },n_{2}^{\prime}}^{(m)
	}\Omega ^{(l)}(t)e^{i(\delta_{m,n_{1},n_{2},n_{1}^{\prime},n_{2}^{\prime}}-\Delta _{l})t}  \notag \\
	&\hspace{0.4cm} \!\times \! \vert m,\!\tilde{n}_{1}(m),\!\tilde{n}_{2}(m)
	\rangle \!\langle m-1,\!\tilde{n}_{1}^{\prime }(m-1),\!\tilde{n%
	}_{2}^{\prime }(m-1) \vert\!+\!\mathrm{H.c.}\Big],
\end{align}}\normalsize
where the detunings are defined as $\delta_{m,n_{1},n_{2},n_{1}^{\prime},n_{2}^{\prime}}=E_{m,n_{1},n_{2}}-E_{m-1,n_{1}^{\prime},n_{2}^{\prime}}=(n_{1}-n_{1}^{\prime})\omega_{M1}+(n_{2}-n_{2}^{\prime})\omega_{M2}-(2m-1)(\chi_{1}+\chi_{2})$.

\subsection{Effective Hamiltonian in a finite-dimensional Hilbert space}
In the above-mentioned work representation, both the cavity mode and the mechanical modes are associated with infinite-dimensional Hilbert spaces. However, under proper parameter conditions, these infinite-dimensional Hilbert spaces can be effectively truncated to finite dimensions. The truncation is guided by resonance transitions induced by pulsed driving fields. Specifically, we choose the carrier frequency $\omega_{l}$ to satisfy the resonance transitions $\vert 0,n_{1},n_{2}\rangle \overset{\Omega^{(1)}(t)}{\longleftrightarrow} \vert 1,\tilde{n}_{1}(1),\tilde{n}_{2}(1)\rangle $ and  $\vert 1,\tilde{n}_{1}(1),\tilde{n}_{2}(1) \rangle \overset{\Omega^{(l\neq1)}(t)}{\longleftrightarrow} \vert 0,n_{1}+N_{l1},n_{2}+N_{l2}\rangle$ with detunings $-\chi_{1}-\chi_{2}$ and $-\chi_{1}-\chi_{2}-N_{l1}\omega_{M1}-N_{l2}\omega_{M2}$, respectively. For the two-mode entangled state we consider, the integers $N_{l1}$ and $N_{l2}$ are restricted to $0$ or $1$. In other scenarios,  $N_{l1}$ and $N_{l2}$ could take different integers depending on the desired states or transitions.

In this case, the Hamiltonian $H_{I}^{(2)}$ can be divided into two parts
\begin{equation}\label{eq21}
H_{I}^{(2)}=\tilde{H}_{I}^{(2)}+\tilde{H}_{I}^{\prime(2)},
\end{equation}
where $\tilde{H}_{I}^{(2)}$ describes the carrier-frequency resonant-transition part
\small{
\begin{align}\label{eq22}
	\tilde{H}_{I}^{(2)}&=\sum_{\substack{n_{1},n_{2}=0}}^{\infty }\Big[A_{n_{1},n_{2},n_{1},n_{2}}^{(1)
	}\Omega ^{(1)}(t)\vert 1,\tilde{n}_{1}(1),\tilde{n}_{2}(1)
	\rangle \langle 0,n_{1},n_{2} \vert \notag \\&\ \hspace{0.4cm} +\sum_{l=2}^{L}A_{n_{1},n_{2},n_{1}+N_{l1},n_{2}+N_{l2}}^{(1)
	}\Omega ^{(l)}(t)\vert 1,\tilde{n}_{1}(1),\tilde{n}_{2}(1)
	\rangle \notag \\ &\ \hspace{0.4cm} \times \langle 0,n_{1}+N_{l1},n_{2}+N_{l2} \vert +\mathrm{H.c.}\Big]
\end{align}}\normalsize
and $\tilde{H}_{I}^{\prime(2)}$ represents the off-resonance transition part
\small{
\begin{align}\label{eq23}
	\tilde{H}_{I}^{\prime(2)}&=\sum_{\substack{m,n_{1},n_{2},n_{1}^{\prime},n_{2}^{\prime}=0}}^{\infty }\,'\Bigg\{A_{n_{1},n_{2},n_{1}^{\prime },n_{2}^{\prime}}^{(m)
	}\Bigg[\Omega ^{(1)}(t)e^{i\delta_{m,n_{1},n_{2},n_{1}^{\prime},n_{2}^{\prime}}^{(1)}t}  \notag \\ & \hspace{0.4cm}+\sum_{l=2}^{L}\Omega ^{(l)}(t) e^{i\delta_{m,n_{1},n_{2},n_{1}^{\prime},n_{2}^{\prime}}^{(l)}t}\Bigg]\vert m,\tilde{n}_{1}(m),\!\tilde{n}_{2}(m)
	\rangle 	\notag \\  &\hspace{0.4cm} \times  \langle m-1,\tilde{n}_{1}^{\prime }(m-1),\!\tilde{n%
	}_{2}^{\prime }(m-1) \vert\!+\!\mathrm{H.c.}\Bigg\}.
\end{align}}\normalsize
Here, the primed summation has the same meaning as in Eq.~(\ref{eq9}) and the detunings for the off-resonant transitions are given by
\small{
\begin{subequations}
	\begin{align}
		\delta_{m,n_{1},n_{2},n_{1}^{\prime},n_{2}^{\prime}}^{(1)} &=(n_{1}-n_{1}^{\prime})\omega_{M1}+(n_{2}-n_{2}^{\prime})\omega_{M2}\notag \\ &\hspace{0.4cm}-2(m-1)(\chi_{1}+\chi_{2}), \label{eq24a}  \\
		\delta_{m,n_{1},n_{2},n_{1}^{\prime},n_{2}^{\prime}}^{(l)} &=(n_{1}-n_{1}^{\prime}+N_{l1})\omega_{M1}+(n_{2}-n_{2}^{\prime}+N_{l2})\omega_{M2}\notag \\ &\hspace{0.4cm}-2(m-1)(\chi_{1}+\chi_{2}) \label{eq24b}.
	\end{align}
\end{subequations}}\normalsize
The conditions $n_{1}\neq n_{1}^{\prime}$ and $n_{2}\neq n_{2}^{\prime}$ in Eq.~(\ref{eq24a}) are satisfied, while $n_{1}^{\prime} \neq n_{1}+N_{l1}$ and $n_{2}^{\prime} \neq n_{2}+N_{l2}$ in Eq.~(\ref{eq24b}) hold when $m=1$.

\begin{figure*}[tbp]
	\centering \includegraphics[width=1\textwidth]{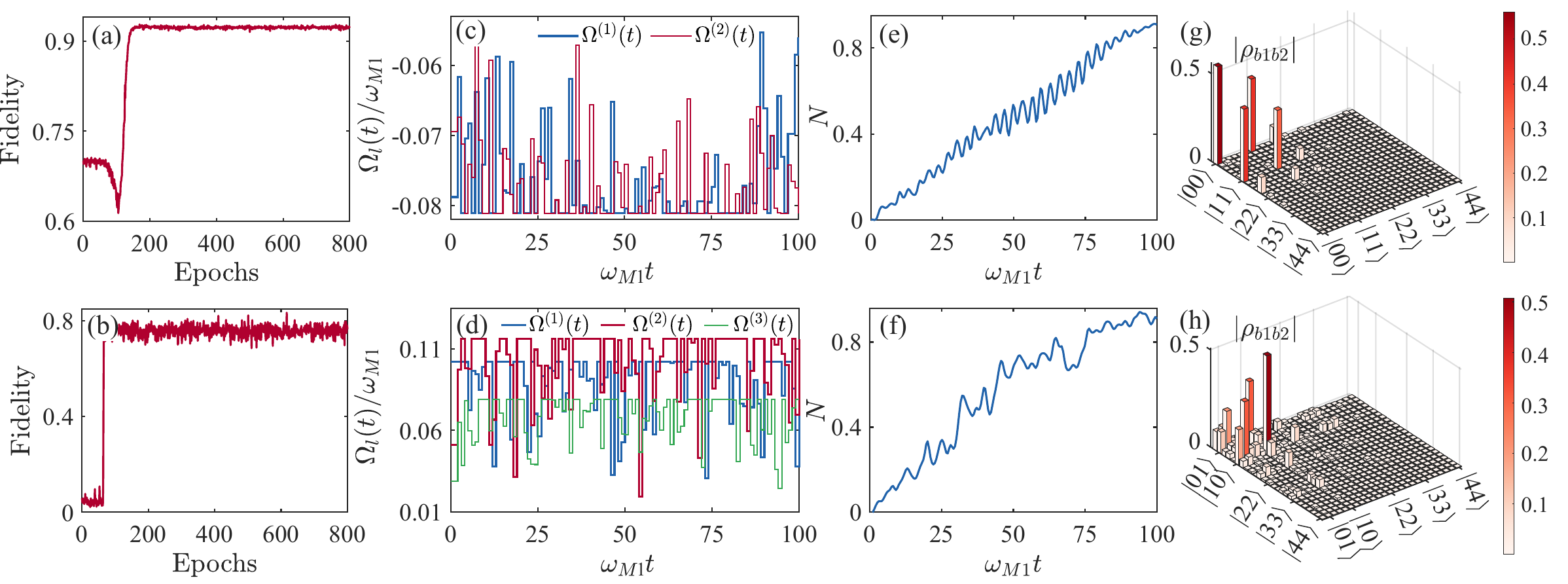}
	\caption{Numerical results for the preparation of the two-mode entangled states. Fidelity is plotted as a function of the training epochs for the states (a) $|\Phi^{(+)}\rangle=(|00\rangle+|11\rangle)/\sqrt{2}$ and (b) $|\Psi^{(+)}\rangle=(|01\rangle+|10\rangle)/\sqrt{2}$. (c) and (d) Optimized time-dependent driving amplitudes $\Omega^{(l)}(t)/\omega_{M1}$, corresponding to the epoch with maximum fidelity in (a) and (b), plotted as a function of $\omega_{M1}t$. (e) and (f) Logarithmic negativity $N$ versus the scaled evolution time $\omega_{M1}t$. (g) and (h) Absolute value of the reduced density-matrix elements in the Fock-state representation for the prepared states. The full Hilbert space consists of 25 basis states, but only the most relevant ones are labeled for clarity. Here we take $N_{c}=3$, $N_{m1}=N_{m2}=5$, $n_{\mathrm{th1}}=n_{\mathrm{th2}}=0$, $g_{01}/\omega_{M1}=1.0$, $S=100$, and $\omega_{M1}T=100$ in all panels. The other parameters are $g_{02}/\omega_{M1}=0.918$, $\omega_{M2}/\omega_{M1}=0.918$, $\kappa/\omega_{M1}=0.002$, and $\gamma_{M1}/\omega_{M1}=\gamma_{M2}/\omega_{M1}=0.0004$ in the top panels and $g_{02}/\omega_{M1}=0.595$, $\omega_{M2}/\omega_{M1}=0.598$, $\kappa/\omega_{M1}=0.001$, and $\gamma_{M1}/\omega_{M1}=\gamma_{M2}/\omega_{M1}=0.0002$ in the bottom panels.}
	\label{Fig6}
\end{figure*}
To suppress the off-resonant transition part $\tilde{H}_{I}^{\prime(2)}$, the parameter conditions $\vert\delta_{m,n_{1},n_{2},n_{1}^{\prime},n_{2}^{\prime}}^{(1,l)}\vert \gg \vert A_{n_{1},n_{2},n_{1}^{\prime },n_{2}^{\prime}}^{(m)}\vert \vert \Omega ^{(1,l)}\vert_{\max}$ should be satisfied, where $\vert \Omega ^{(1,l)}(t)\vert_{\max}$ represents the maximum amplitude of the pulsed driving fields. Specifically, to minimize the transitions between the states $\vert 1,\tilde{n}_{1}(1),\tilde{n}_{2}(1) \rangle$ and  $\vert 2,\tilde{n}_{1}(2),\tilde{n}_{2}(2) \rangle$, the condition
\begin{equation}\label{eq25}
	\vert\delta_{2,n_{1},n_{2},n_{1}^{\prime},n_{2}^{\prime}}^{(1,l)}\vert \gg \vert A_{n_{1},n_{2},n_{1}^{\prime },n_{2}^{\prime}}^{(2)}\vert \vert \Omega ^{(1,l)}\vert_{\max}
\end{equation}
should be satisfied. Under these conditions, the Hamiltonian $H_{I}^{(2)}$ can be simplified to $\tilde{H}_{I}^{(2)}$. Using a similar truncation used in the single-resonator case, the effective Hamiltonian now becomes
\begin{align}\label{eq26}
	\tilde{H}_{I}^{(2)}&=\sum_{\substack{n_{1},n_{2}=0}}^{N_{m}-1 }A_{n_{1},n_{2},n_{1},n_{2}}^{(1)
	}\Omega ^{(1)}(t)\vert 1,\tilde{n}_{1}(1),\tilde{n}_{2}(1)
	\rangle \langle 0,n_{1},n_{2} \vert \notag \\&\ \hspace{0.4cm} +\sum_{l=2}^{L}\sum_{\substack{n_{1}=0}}^{N_{m}-N_{l1} }\sum_{\substack{n_{2}=0}}^{N_{m}-N_{l2} }A_{n_{1},n_{2},n_{1}+N_{l1},n_{2}+N_{l2}}^{(1)
	}\Omega ^{(l)}(t) \notag \\ &\ \hspace{0.4cm} \times \vert 1,\tilde{n}_{1}(1),\tilde{s}(1)
	\rangle\langle 0,n_{1}\!+\!N_{l1},n_{2}\!+\!N_{l2} \vert +\mathrm{H.c.}.
\end{align}
We assume that the initial state of the system is $\vert 0,0,0 \rangle$. The transitions $\vert 0,1,1\rangle \overset{\Omega^{(1)}(t)}{\longleftrightarrow} \vert 1,\tilde{1}(1),\tilde{1}(1)\rangle $, $\vert 0,0,1\rangle \overset{\Omega^{(1)}(t)}{\longleftrightarrow} \vert 1,\tilde{0}(1),\tilde{1}(1)\rangle $, and $\vert 0,1,0\rangle \overset{\Omega^{(1)}(t)}{\longleftrightarrow} \vert 1,\tilde{1}(1),\tilde{0}(1)\rangle $ are negligible because the states $\vert 0,1,1\rangle$, $\vert 0,0,1\rangle$, and $\vert 0,1,0\rangle$ are initially unpopulated, and the conditions $A_{N_{l1},N_{l2},N_{l1},N_{l2}}^{(1)}\approx0$ further suppress these transitions. In contrast, the transition $\vert 0,0,0 \rangle\overset{\Omega^{(1)}(t)}{\longleftrightarrow} \vert 1,\tilde{0}(1),\tilde{0}(1)\rangle$ plays a dominant role.

\subsection{Generation of entangled states of the two mechanical resonators}
To exhibit the validity of the method, we consider the generation of two-mode entangled states: $|\Phi^{(\pm)}\rangle=(|00\rangle\pm|11\rangle)/\sqrt{2}$ and $|\Psi^{(\pm)}\rangle=(|01\rangle\pm|10\rangle)/\sqrt{2}$. Concretely, to generate $|\Phi^{(+)}\rangle$, two pulsed driving fields are applied, with the carrier frequencies satisfying the detunings $\Delta_{1}=-\chi_{1}-\chi_{2}$ and $\Delta_{2} = -\chi_{1}-\chi_{2}-\omega_{M1}-\omega_{M2}$, respectively. Differently, for the generation of the state $|\Psi^{(+)}\rangle$, three pulsed driving fields are required, with carrier frequencies satisfying the detunings $\Delta_{1}=-\chi_{1}-\chi_{2}$, $\Delta_{2} = -\chi_{1}-\chi_{2}-\omega_{M1}$, and $\Delta_{3} = -\chi_{1}-\chi_{2}-\omega_{M2}$.
	
To describe the system dynamics of the two-resonator optomechanical system, we generalize the dressed master equation~(\ref{eq13}) to the two-resonator case. The quantum master equation for the two-resonator optomechanical system is given by
{\small
	\begin{align}
	\!\dot{\rho}^{(2)}(t)&\! =\!-i[H_{R}^{(2)},\rho ^{(2)}(t)]\!+\!\sum_{i=1,2}\{\gamma
	_{Mi}(n_{\mathrm{thi}}+1)\mathcal{D}[b_{i}-\beta _{i}a^{\dagger }a]  \nonumber \\
	& ~~~\!\times\! \rho ^{(2)}(t)+\gamma
	_{Mi}n_{\mathrm{thi}}\mathcal{D}[b_{i}^{\dagger }-\beta _{i}a^{\dagger
	}a]\rho ^{(2)}(t)+4\gamma _{Mi}  \nonumber \\
	& ~~~\!\times\!(k_{B}T_{bi}/\omega _{Mi}) \beta
	_{i}^{2}\mathcal{D}[a^{\dagger }a]\rho ^{(2)}(t)\}\!+\!\kappa \mathcal{D}[a]\rho
	^{(2)}(t),
\end{align}}
where $\rho ^{(2)}(t)$ is the density matrix of the two-resonator optomechanical system and $H_{R}^{(2)}$ is defined in Eq.~(\ref{eq15}). The parameter $\gamma_{Mi}$ denotes the decay rate of the $i$th mechanical mode and $n_{\mathrm{th}i}$ is the thermal phonon occupation for $i$th mechanical mode at temperature $T_{bi}$. Similar to the state preparation in the single-resonator system, the fidelity $F_{b}$ between the target states and the numerically simulated results is maximized by defining a reward function $r_{t}=-10\mathrm{log}_{10}(1-F_{b})$. The optimization of the pulsed driving fields to achieve the desired entangled states is guided by the DDPG algorithm.

In Fig.~\ref{Fig6} we display the optimization and characterization of the two-mode entangled states $|\Phi^{(+)}\rangle$ and $|\Psi^{(+)}\rangle$. Concretely, we show the fidelity as a function of the training epochs corresponding to the preparation of $|\Phi^{(+)}\rangle$ and $|\Psi^{(+)}\rangle$ in Figs.~\ref{Fig6}(a) and \ref{Fig6}(b), respectively. During the early stages of the training, the fidelity improves slowly, reflecting the exploration of the parameter space by the DDPG algorithm. The maximum fidelities of $0.928$ and $0.834$ are achieved at epochs $691$ and $617$ for the preparation of $|\Phi^{(+)}\rangle$ and $|\Psi^{(+)}\rangle$, respectively. The results corresponding to these maximum fidelities are further analyzed in Figs.~\ref{Fig6}(c),~\ref{Fig6}(e), and~\ref{Fig6}(g) for $|\Phi^{(+)}\rangle$ and in Figs.~\ref{Fig6}(d),~\ref{Fig6}(f), and~\ref{Fig6}(h) for $|\Psi^{(+)}\rangle$. Figures~\ref{Fig6}(c) and \ref{Fig6}(d) show the optimized driving amplitudes $\Omega^{(l)}(t)/\omega_{M1}$ versus the scaled evolution time $\omega_{M1}t$. The optimized driving fields exhibit complex time-varying structures. In these simulations, we checked that the condition in Eq.~(\ref{eq25}) can be well satisfied.

To quantify the degree of entanglement in the generated states, we adopt the logarithmic negativity as the entanglement measure. For a two-mode system described by the density matrix $\rho_{b1b2}$, the logarithmic negativity is defined by $N=\log _{2}\Vert \rho _{b1b2}^{T_{b2}}\Vert _{1}$~\cite{PlenioPRL2005}, where $T_{b2}$ denotes the partial transpose of the density matrix $\rho_{b1b2}$ with respect to the mechanical mode $b_{2}$, and the trace norm is defined as $\Vert \rho _{b1b2}^{T_{b2}}
\Vert _{1}=\mathrm{Tr}[\sqrt{(\rho _{b1b2}^{T_{b2}}) ^{\dag }\rho _{b1b2}^{T_{b2}}}]$. The time evolution of the logarithmic negativity for the two generated states is presented in Figs.~\ref{Fig6}(e) and \ref{Fig6}(f). It can be observed that the logarithmic negativity of the two generated states gradually increases over time, exhibiting small oscillations.  At the target time, the logarithmic negativity reaches approximately $0.91$, indicating an expected quantum entanglement between the two mechanical modes. It is worth noting that our reinforcement learning framework is explicitly designed to maximize the state fidelity at the predetermined time $T$, rather than to continuously optimize entanglement throughout the entire evolution. As a result, the logarithmic negativity at time $T$ represents a balance between entanglement generation and the unavoidable effects of dissipation accumulated during the evolution process.

To illustrate the quantum state of the system, Figs.~\ref{Fig6}(g) and \ref{Fig6}(h) show the absolute values of the reduced density-matrix elements in the Fock-state representation of $\rho_{b1b2}$, which is obtained by tracing out the cavity mode from the system density matrix. For both generated states, the off-diagonal elements indicate quantum coherence between the mechanical modes, while the diagonal elements reveal the expected population distributions within the finite-dimensional Hilbert space. These results confirm the successful preparation of the desired two-mode entangled states and provide a comprehensive characterization of their quantum properties.

\section{Discussions and Conclusion\label{Sec4}}
We now discuss the experimental feasibility of the proposed scheme. Implementing the state preparation requires two key conditions: single-photon ultrastrong optomechanical coupling and complex square-wave driving fields. Concerning the coupling regime, we point out that the single-photon ultrastrong coupling has not been realized in experiments. However, many recent advances have been made in the enhancement of single-photon optomechanical coupling. For example, in superconducting circuits, coupling strengths of $g_{0}/2\pi=10$ MHz~\cite{Massel2014} and $g_{0}/2\pi=1.6$ MHz~\cite{MasselNC2015} have been reported, with the potential to improve up to $g_{0}/2\pi=100$~MHz through further device optimization~\cite{MasselNC2015}. Other relevant parameters from Ref.~\cite{MasselNC2015} are $\omega_{M}/2\pi\simeq65$~MHz, $\gamma_{M}/2\pi =15$ kHz, and $\kappa/2\pi\simeq 1.27$~MHz. In our numerical simulations, the single-photon optomechanical coupling strength employed exceeds currently available experimental values. However, the cavity and mechanical decay rates are chosen within experimentally achievable ranges, suggesting that the scheme could become achievable with future improvements in coupling strength. To satisfy the second requirement, the generation of a square-wave driving field is essential. While square-wave driving has already been realized in superconducting-circuit platforms~\cite{WallraffPRL2005}, implementing more complex pulse sequences remains challenging due to the fast variations required in this scheme. However, ongoing technological advancements are expected to overcome these difficulties, making the generation of such complex square-wave driving fields feasible in future experiments.

We further compare our scheme with conventional methods, such as the STIRAP~\cite{VitanovRMP2017}, which has also been applied to the preparation of phononic Fock states~\cite{ZouPRA2022}. In that work, a phononic Fock state with $N=2$ was generated via STIRAP, and the emission dynamics of correlated phonon pairs was analyzed. By appropriately tuning the resonance conditions, STIRAP can in principle be extended to prepare larger phononic Fock states. Moreover, superposition states of the form $(|0\rangle+|N\rangle)\sqrt{2}$ can be obtained by selecting suitable time points during the adiabatic evolution. However, the STIRAP technique faces two intrinsic limitations. On the one hand, it requires long evolution times to satisfy the adiabatic condition, which can reduce the fidelity due to dissipation. On the other hand, it becomes inefficient when preparing target states that involve multiple driving pulses and complex parameter tuning, such as the superposition state $(|1\rangle + |2\rangle)/\sqrt{2}$ and the entangled state $|\Psi^{(\pm)}\rangle = (|01\rangle \pm |10\rangle)/\sqrt{2}$ considered in this work. In contrast, our reinforcement learning-based approach enables high-fidelity state preparation within shorter evolution times and offers greater flexibility in controlling multiple driving parameters, making it more suitable for the generation of complex non-classical mechanical states.

Building on the previous discussion, the quantum states generated in our scheme, including phononic Fock states, superposed Fock states, and two-mode entangled states, can be detected using high-resolution single-photon spectroscopy. In the resolved-sideband regime, where the mechanical frequency significantly exceeds the cavity linewidth $\kappa$, phonon sidebands appearing in the output optical spectrum encode information about the phonon number distribution and displacement characteristics of the mechanical modes, thereby enabling indirect identification of the prepared quantum states~\cite{LiaoPRASpe2012}. For two-mode entangled states, additional spectral signatures may reflect the presence of nonclassical correlations. However, complete verification of entanglement typically requires joint measurements or the application of entanglement witnesses.

In conclusion, we have proposed a scheme for generating several nonclassical mechanical states in cavity optomechanical systems working in the single-photon ultrastrong-coupling regime. By analyzing the resonance conditions in the eigenstate representation and optimizing the pulsed driving amplitudes using the DDPG algorithm, we have successfully proved the preparation of these nonclassical mechanical states. In particular, in the presence of the system dissipations, we have realized the high-fidelity generation of phononic Fock states and superposed Fock states in the single-resonator optomechanical system. We have also achieved the high-fidelity preparation of two-mode entangled states in the two-resonator optomechanical system. Our work creates an opportunity for manipulating cavity optomechanical systems based on the reinforcement learning method.

\begin{acknowledgments}
Y.-H.L. thanks Dr.~Fen Zou and Mr.~Ying Hu for helpful discussions. J.-Q.L. was supported in part by National Natural Science Foundation of China (Grants No.~12175061, No.~12247105, No.~11935006, and No.~12421005), National Key Research and Development Program of China (Grant No.~2024YFE0102400), and Hunan Provincial Major Sci-Tech Program (Grant No.~2023ZJ1010). Q.-S.T. was supported in part by the National Natural Science Foundation of China (Grant No.~12275077) and the Natural Science Foundation of Hunan Province (Grant
No.~2022JJ30277). L.-M.K. was supported by the Natural Science Foundation of China (Grants No. 12247105, No. 12175060 and No. 12421005), the Hunan Provincial Major Sci-Tech Program (Grant No. 2023ZJ1010), and the XJ-Lab key project (Grant No. 23XJ02001).
\end{acknowledgments}

\bibliography{ref}

\end{document}